# Bulk and Lattice Properties for Rigid Carbon Nanotubes Materials

V.K. Jindal[1], Shuchi Gupta and K. Dharamvir

Department of Physics, Panjab University, Chandigarh-160014, India

We use an atom-atom potential between carbon atoms to obtain an interaction potential between nanotubes (assumed rigid), thereby calculating the cohesive energy of a bunch of nanotubes in hexagonal two dimensional packing. The model proposed is quite similar to our earlier work on fullerenes and organic molecular crystals. The results for inter-nanotube distances, energy per unit length, bulk modulus and phonons for inter-nanotube vibrations are obtained and compared with available data from measurements and other available calculations. We also model formation of multi-wall nanotubes. We find the results for various calculated quantities agreeing very well with measured structural parameters and other calculations. The reversible energy stored on compression of the bunch of nanotubes on application of pressure up to 30 Kbar calculated in this rigid molecule model is overestimated by about 30% when compared with measured results, signifying the appreciable flexibility of tubes at high pressures. The model is considered very suitable for incorporating flexible nanotubes in bunches of single and multi-wall nanotube materials of various types.

**I. Introduction:**

Followed by the discovery of fullerene molecules[1] comprising of 60 or 70 or larger number of carbon atoms bonded in round or nearly round cage-like shapes, and subsequent synthesization of solids[2] of these molecules in the laboratory in early 90's, tremendous scientific effort has been made to understand, interpret and hypothesize properties of these carbon based solids. The unique carbon based materials have appeared in cluster forms like $C_{60}$, $C_{70}$ solids, their aggregates and carbon nanotubes. Solid $C_{60}$ is the most extensively studied fullerene solid. Sufficient results on properties of these solids based on structure, lattice dynamics, pressure and temperature effects, optical as well as transport and thermal properties are now available experimentally and some interpreted theoretically. In oversimplified terms, one can safely assume the intermolecular interactions to be governed by Van-der-Waals like interactions, which are generally derived by summing carbon atom – atom interactions of the involved molecules [3-7]. Similarly, polymerized fullerene solids have also been obtained[8,9] by various processes like photo-polymerization, electron beam induced polymerization, pressure induced polymerization and plasma-induced polymerization processes. Similar theoretical models for dimerized and polymerized solids have been proposed[10]. Another class of fullerene solids have been obtained by doping pure fullerene solids. The doping may be obtained with dopant inside the cage (endohedral doping), or on the shell (substitutional doping) or exohedral doping where dopant is outside the shell on the lattice. The exhohedral doping in which the dopants are alkali metals have shown interesting superconducting properties. Therefore, much attention has also been paid to understanding the electronic properties of $C_{60}$ molecules and solids derived from them[11-15]. A useful information about the experimental data and theoretical models for $C_{60}$ molecules and their solids can be found in some of the reviews which have been published from time to time, e.g., Copley et. al.[16], Ramirez[17] and Dresselhaus[18].

---

[1] The author, with whom correspondence be made. E-mail: *jindal@panjabuniv.chd.nic.in*

The clusters of carbon atoms, in addition to round or nearly round shapes, have also been synthesized in cylindrical or nearly cylindrical forms, called nanotubes. These nanotubes are very close to one-dimensional crystals, as the ratio of length to diameter can be grown to be $\sim 10^6$. Carbon nanotubes were discovered in 1991 by S. Iijima[19]. These are large macromolecules that are unique for their size, shape, and remarkable physical properties. They can be thought of as a sheet of graphite (a hexagonal lattice of carbon) rolled into a cylinder to form a macro-molecule of carbon . These intriguing structures have sparked much excitement in the recent years as the inside of the nanotube is extremely fascinating, being hollow, a candidate for trap for various sized atoms, and for study of quantum phenomena prevalent at dimensions of its diameter. Indeed, practical applications, including hydrogen uptake, has also generated research interest dedicated to their understanding. Currently, the physical properties are still being discovered and disputed. Interestingly, nanotubes have a very broad range of electronic, thermal, and structural properties that change depending on the different kinds of nanotube (defined by its diameter, length, and chirality, or twist). Besides having a single cylindrical wall (SWNTs), nanotubes can have multiple walls (MWNTs)--cylinders inside the other cylinders. Nanotubes have been known to be up to one hundred times as strong as steel and almost two millimeters long. These nanotubes may have a hemispherical "cap" at each end of the cylinder. They are light, flexible, thermally stable, and are chemically inert. They have the ability to be either metallic or semi-conducting depending on the "twist" of the tube. Infact, they are identified as "armchair" nanotube,"zigzag" type or "chiral" type depending upon folding of graphite sheet, indicated in literature by two integers, (n,m), dictating which atom to join with on folding the graphite sheet. The chirality in turn affects the conductance of the nanotube, it's density, it's lattice structure, and other properties. A nice description about fullerene nanotubes has been given by Yakobson and Smalley[20]. It turns out that the average diameter of a SWNT is 1.4 nm. However, nanotubes can vary in size, and they aren't always perfectly cylindrical. The carbon bond length of 0.142 nm was measured[21].

The carbon nanotube materials in the form of 'ropes' have been extensively studied by Thess et al.[22] Ropes are bundles of tubes packed together in an orderly manner. They found that the individual SWNTs packed into a close-packed triangular lattice with a lattice constant of about 17 Å, though the actual value depends on whether the material consists of armchair, zigzag or other chirality. Gao et al[23] have presented extensive molecular dynamics results , using various cross sections of SWNT, for structural and mechanical properties. Armchair tubes had a lattice parameter of 16.78 Å and had a density of 1.33 g/cm$^3$. For MWNT materials, the inter-wall separation between the tubes was also dependent on chirality. Armchair tubes had a spacing of 3.38 Å, zigzag tubes had a spacing of 3.41 Å, and (2n, n) chiral tubes had interlayer spacing value of 3.39 Å. These values of spacing compare with the spacing between the layers of graphite sheets, both are about 3.4 Å. The maximum tensile strength was found close to 30 Gpa[24]. However, there has been some controversy into the value of the modulus perhaps due to interpretation of the thickness of the walls of the nanotube. If the tube is considered to be a solid cylinder, then it would have a lower Young's modulus. If the tube is considered to be hollow, the modulus gets higher, and the thinner we treat the walls of the nanotube, the higher the modulus will become.

Tersoff and Ruoff[25] have studied structural properties of bundles of carbon nanotubes, and presented detailed calculated results assuming van-der Waals interactions between cylindrical and deformed cylinder of various diameters, for lattice constant and cohesive energy. Recently[26], Chesnokov et. al. have measured compression of nanotube materials under pressure and find complete reversible results for density to restore back upon release of pressure upto 29Kbar. A van-der Waals interaction based model for nanotube bundles has also been proposed[27] recently by Henrad et. al., for a continuum model.

In this paper we attempt to calculate detailed lattice dynamical properties of materials of carbon nanotubes of various sizes and types forming solids either as bunches in hexagonal packing or MWNT formations. Our aim has been to suggest a simple model that explains the mechanical properties of nanotube materials. The success of this model is found to lead us to include intra-molecular interactions especially for bending modes to deal with high pressure effects on flexible nanotube materials. This part of the problem is progressing and would be published separately[28]. Effect of hydrostatic pressure has been studied on bulk, structure and phonon related properties.

We give general details regarding the theoretical procedure, necessary expressions etc. in section II. We also give in this section the numerical procedure leading to calculations of various properties concerning the subject matter. The results are compared with available experimental data and other results, and discussed in section III. Finally, the findings of this paper are summarized and concluded in section IV.

## II. Generalities

In order to calculate various bulk, structural and lattice dynamical properties in materials of nanotubes, we use the procedure and approach as adopted earlier by us for pure $C_{60}$ solids. We assume rigid nanotubes in the present work, and that the inter-nanotube potential energy is a sum of atom-atom interactions. The position co-ordinates of carbon atoms on the surface of a nanotube can be easily determined for a given (n,m) nanotube of varying length. In our calculation, we take various lengths to ascertain the length of the nanotube that can be considered long enough for the results obtained for various calculated quantities per unit length to be independent of the length. We therefore propose a long tube model.

## A. Co-ordinates of C atoms

Long part of a nanotube (excluding the caps) can be constructed by rolling up a graphene sheet into a cylinder. The rolling up can be made in various ways, involving integer multiples of basic vectors of hexagons forming the graphite sheet. If these two integers are denoted by n and m, the nanotube is usually denoted as (n, m) nanotube, and m can be 0, equal to n or any other integer. The nanotube is accordingly named as armchair, zigzag or a chiral one. A typical armchair nanotube is shown in fig. 1. Knowing n and m determines the geometry of the nanotube and hence the radius gets defined and all atom coordinates for any length, (multiple of some basic length) can be generated. In our model, we use these discrete atom positions rather than the uniform cylindrical model for our calculation. Therefore, in this model it is possible to keep distinction between different types of nanotubes. With M as the total number of carbon atoms on a nanotube (n,m) with hexagonal bond length as $a_h$, the diameter d and length l of the tube are given as:

$d = \sqrt{3}\, a_h\, (m^2 + n^2 + mn)^{1/2} / \pi$

$l = M\sqrt{3}\, a_h / 4n$      (for armchair and zigzag tubes)

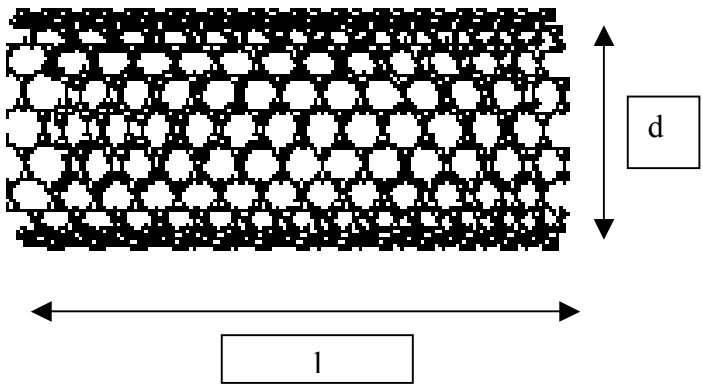

Fig.1. A typical armchair nanotube of diameter d and length l.

## B. Crystalline Structure

The long nanotubes, assumed rigid tend to bunch as 2-D hexagonal packing, with each central nanotube surrounded by 6 other nanotubes. A typical 2-D structure and unit cell is shown in fig. 2.

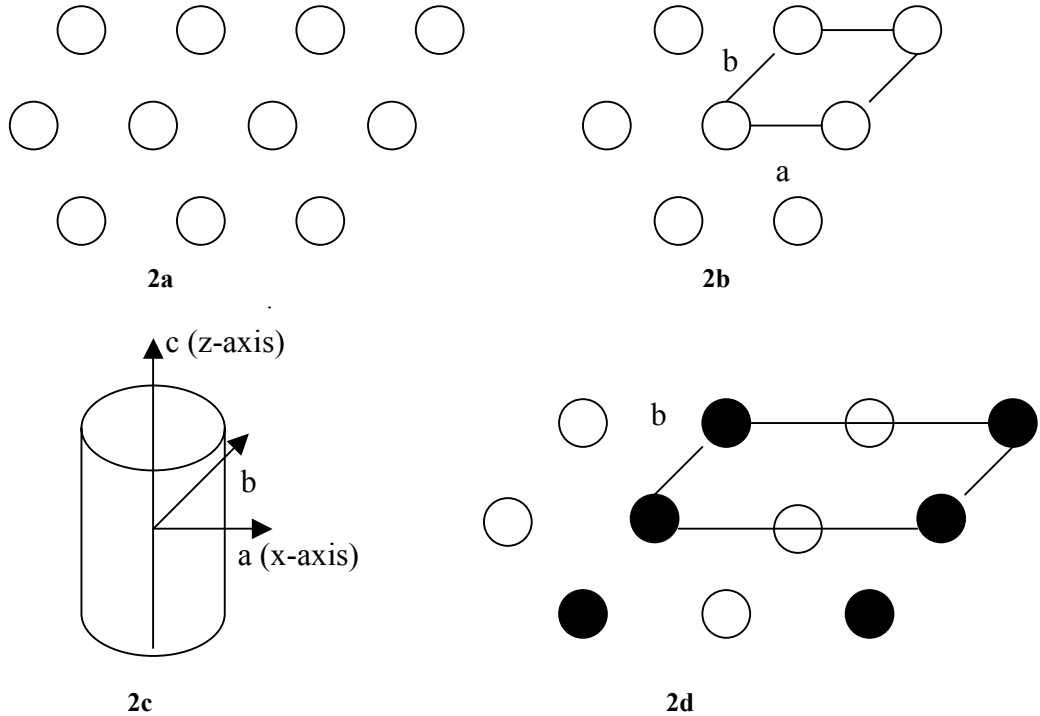

Fig 2. The figure a shows 2-D hexagonal structure of the nanotube crystal. 2b represents the mono molecular unit cell. 2c shows a single isolated nanotube and Co-ordinates system. 2d shows the di-molecular unit cell chosen with 2 molecules, shown as shaded and unshaded, which differ in orientation along 'c' axis by $180^o$.

The two-dimensional position vector $\mathbf{R_l}$ of any nanotube center on the two-dimensional lattice is given by

$\mathbf{R_l}$= l**a**+m**b**, where l and m are any two integers.

## C. Model Potential

The inter-tube potential energy $U_{l\kappa,l'\kappa'}$ between two nanotubes (molecules), identified by $\kappa$ molecule in unit cell index **l**, $\kappa'$ molecule in cell **l'**, can be written as a pair-wise sum of C-atom-atom potentials (C-C) on these two molecules, i.e.

$$U_{l\kappa,l'\kappa'} = \sum_{ij} V(r_{ij}),  \quad (1)$$

where the sum in Eq. 1 includes all the M atoms in each of the nanotube molecules, and V(r) is the C-C potential. We take the potential V(r), where r is the distance between the C-C atoms, to be given by

$$V(r) = -A/r^6 + B\exp(-\alpha r) \quad (2)$$

The interaction parameters A, B and $\alpha$ have been obtained for various atom-atom interactions from gas phase data, and we obtain these for our use from the set provided by Kitaigorodski[29]. For C-C interaction, these parameters have been tabulated in Table I.

### Table I
Atom-atom potential parameters (Kitaigorodski[29])

| A=358 kcal/mole-$A^6$ | B=42000 kcal/mole | $\alpha$=3.58$A^{-1}$ |
|---|---|---|

The total potential energy $\Phi$ can be obtained by carrying out the lattice sum, knowing the position of the lattice points,

$$\Phi = 1/2 \sum_{l\kappa,l'\kappa'}' U_{l\kappa,l'\kappa'} \quad (3)$$

**D. Cohesive Energy**

In order to calculate the cohesive energy for the nanotube material with long 2-D hexagonal packing, the intermolecular potential energy (Eq. 3) needs to be obtained. The lattice parameter and orientation angle ($\phi$) of a nanotube along the long nanotube axis (c-axis) with respect to a given initial orientation are varied till a minimum is attained in the total potential energy. The summation over various lattice points in Eq. 3 is carried out numerically for various intermolecular distances and asymptotic values of the total energy, for extended 2-D crystalline bunch are obtained.

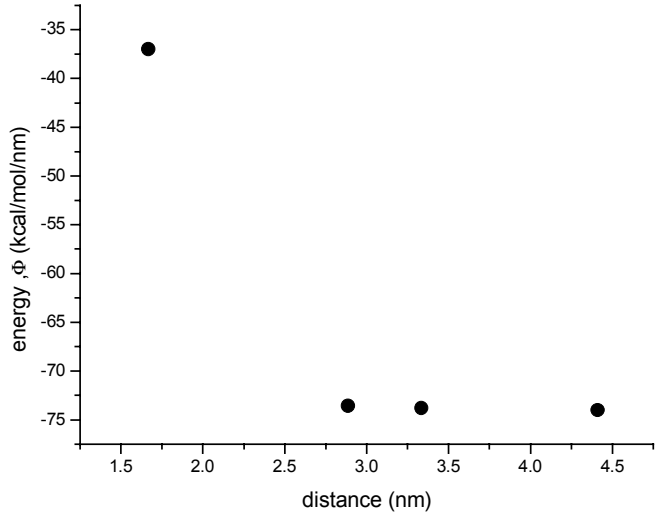

Fig 3. The calculated potential energy of the SWNT crystal (Eq.2) by restricting summation upto various distances from the nanotube at the origin.

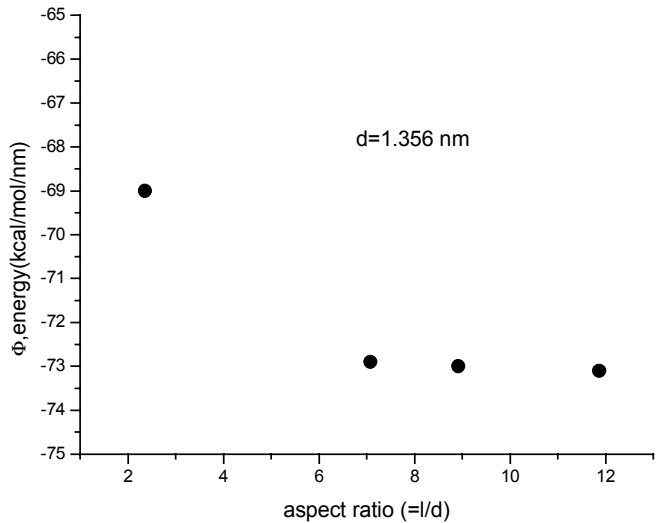

Fig. 4. Potential energy of the nanotube crystal per unit length calculated for various lengths of the nanotube.

The potential energy obtained in this way is a function of the lattice parameter and orientations of the molecules in the unit cell. In order to check the convergence of the lattice sums, the potential energy calculation was made by including in the summation, various neighbours, (Fig. 3) and an estimate of the asymptotic value of energy was obtained. In view of the results of potential energy as a function of lattice distance as shown in Fig. 3, summation upto second nearest neighbours corresponding to 2.8 nm is sufficient. Similarly, the length of the nanotubes was varied to obtain the potential energy per unit length. Finally, a length to diameter ratio of 8 was considered to be a good approximation for calculated potential energy per unit length to be independent of the length of the tube The results for the potential energy as a function of aspect ratio is shown in fig. 4. Indeed it was found that an aspect ratio of 8 was good enough for treating the tubes as infinitely long when the calculation was repeated for various other diameters of the nanotubes as well.

A minimization of the potential energy is necessary to obtain the equilibrium orientations and lattice parameter. Numerical results for lattice parameter (a) and total potential energy ($\Phi$) thus obtained in minimised energy configuration are presented in Table IIA, for armchair and zigzag nanotubes. For comparison with other carbon clusters , the potential energy/C-atom has also been tabulated in Table IIB. The plots of potential energy as a function of lattice parameter as well as that of the angle of orientation $\phi$ are presented in Figs. 5 and 6, respectively, for armchair nanotube bunch.

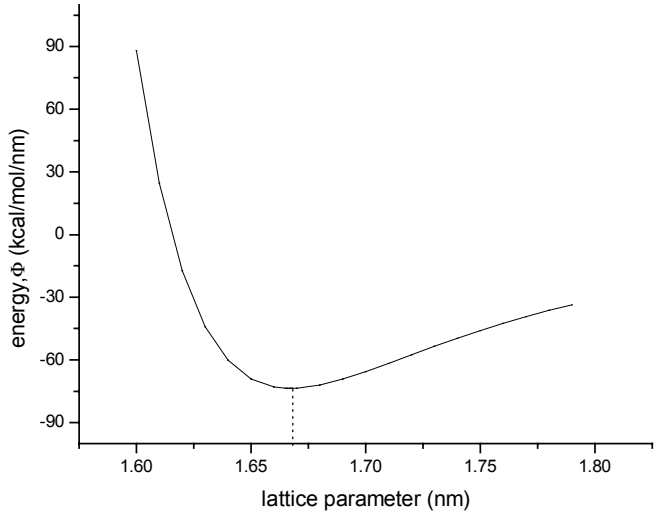

Fig. 5. Calculated potential energy per unit length as a function of the lattice parameter. Equilibrium value of the lattice constant comes out to be 1.667 nm.

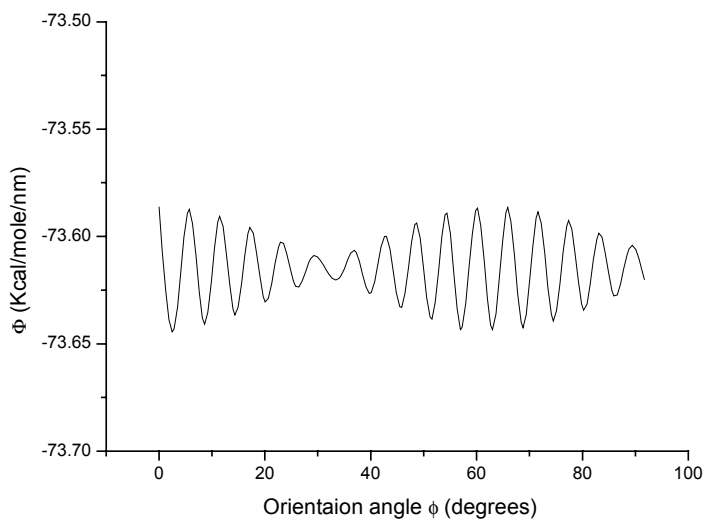

Fig 6  Potential energy calculation using various orientations of the nanotubes along z-axis.

**Table II A**

Potential Energy and Structure of SWNT Crystals

| Type of the tube | Diameter (in nm) | Lattice parameter (in nm) | | Total potential energy/length (kcal/mol/nm) | Other calculation[25] (kcal/mol/nm) |
|---|---|---|---|---|---|
| | | Present | Expt.[21] | | |
| Armchair (10,10) | 1.356 | 1.667 | 1.678 | 72.9 | 63.6 |
| Zig-zag (17,0) | 1.332 | 1.642 | 1.652 | 72.7 | |

**Table II B**

Comparison of Potential Energy of various Carbon Cluster Solids

| Type of the cluster | Potential energy/ C-atom (meV/C-atom) |
|---|---|
| Armchair SWNT Crystal | 19.33 |
| $C_{60}$ Crystal[7] | 29.14 |
| $C_{70}$ Crystal[30] | 23.98 |

**E. Free-Rotation Model**

For the case of freely rotating molecules, (i.e. near and above room temperature), the nanotube molecules can be replaced by cylindrical shells, with uniform surface density of carbon atoms. In this way, Eqs. 1 and 2 give $U_{ll}$, as interaction energy per unit length as

$$U_{ll}(R) = -U_{attractive}(R) + U_{repulsive}(R), \quad (4)$$

where

$$U_{attractive}(R) = \frac{8}{3}\pi^3 \sigma^2 A \frac{r^2}{R^5} \sum_{n=0}^{\infty} \left\{ {}^{2n}C_n \left(n+\tfrac{1}{2}\right)\left(n+\tfrac{3}{2}\right) \right\}^2 \left(\frac{r}{2R}\right)^{2n} \quad (5)$$

and

$$U_{repulsive}(R) = 8\pi^2 \sigma^2 B \{rI_0(\alpha r)\}^2 RK_1(\alpha r) \left[1 - \frac{2r}{R}\frac{I_1(\alpha r)}{I_0(\alpha r)}\frac{K_0(\alpha R)}{K_1(\alpha R)}\right] \quad (6)$$

with r as the radius of a nano-tube, R representing the distance between the axes of the two tubes which are parallel to each other and σ is the number density of C-atoms on its surface, i.e. 2 per hexagon and equals $38.17/nm^2$. $I_n$ is the modified Bessel function of the first kind of order n and $K_n$ is the modified Bessel function of the second kind of order n. ${}^{2n}C_n$ are the Binomial coefficients.

**F. MWNT Materials**

A multi-wall nanotube can be formed by concentric layering of nanotubes of ever increasing diameters. We have calculated the configuration of these using a model similar to that for SWNT material for armchair nanotubes. The inter-wall separation obtained in this model in minimum energy configuration of 2-wall and 3-Wall nanotube is presented in Table III

**Table III**

**Energy and Structure of layered nanotubes (MWNT)**

| MWNT Layer | Diameter (nm) | Interwall separation (in nm) | | Energy /length (kcal/mol/nm) |
| --- | --- | --- | --- | --- |
| | | Present | Expt.[21] | |
| 1$^{st}$ tube | 1.35 | - | - | 72.9 |
| 2$^{nd}$ tube | 2.03 | 0.339 | 0.338 | 211.6 |
| 3$^{rd}$ tube | 2.71 | 0.339 | 0.338 | 535.2 |

**G. Harmonic Phonons**

The nanotube material is a molecular crystal, having vibrational and librational modes. The total potential energy of the crystal is dependent upon the intermolecular separation as well as on the orientation of the molecules. For the sake of convenience and generality with 3-D systems, we describe its lattice dynamics also by a Taylor series expansion in terms of a six-component translation-rotation displacement vector $u_\mu(l\kappa)$ ( instead of that suitable for a 2-D system) representing $\mu$ th component of the displacement of $\kappa$ th molecule in **l** th cell. Here the index $\mu$ runs from 1 to 6, the first three components representing the translational displacement (the x, y and z components) and the other three, the rotational displacements as angles about the three axes. The three axes are profitably chosen as the principal axes of moment of inertia of a molecule, as this simplifies the expressions for the crystal Hamiltonian. The expression for the molecular crystal Hamiltonian is then written as[31]

$$H = \frac{1}{2}\sum_{l\kappa\mu} m_\mu(\kappa)\left[\dot{u}_\mu(l\kappa)\right]^2 + \frac{1}{2}\sum_{l_1 l_2}\sum_{\kappa_1\kappa_2}\sum_{\mu_1\mu_2}\Phi_{\mu_1\mu_2}(l_1\kappa_1,l_2\kappa_2)u_{\mu_1}(l_1\kappa_2)u_{\mu_2}(l_2\kappa_2), \qquad (6)$$

where we retain only the harmonic part of the potential energy. The kinetic energy part involves the translational energy for $\mu \leq 3$, when $m_\mu(\kappa)$ represents the mass of the molecule at $\kappa$, which in present case is independent of $\kappa$. For $\mu > 3$, $m_\mu(\kappa)$ represents the moment of inertia along the principal axes for the $\kappa$th molecule and kinetic energy corresponds to rotational kinetic energy. $\Phi_{\mu_1\mu_2}(l_1\kappa,l_2\kappa_2)$ is a harmonic force constant, and is defined in terms of second derivative of the potential energy at equilibrium, i.e.

$$\Phi_{\mu_1\mu_2}(l_1\kappa,l_2\kappa_2)_1 = \left(\frac{\partial^2 \Phi}{\partial u_{\mu_1}(l_1\kappa_1)\partial u_{\mu_2}(l_2\kappa_2)}\right)_0 \qquad (7)$$

Since $\Phi$ is related to atom-atom potential V(r), (Eqs. 1-3), the force constant in Eq. (7) can be expressed in terms of V(r), with appropriate transformation relations involving the relationship of the atom-atom distance r as functions of molecule translation or rotations. Finally, the dynamical matrix defined as,

$$M_{\mu_1\mu_2}(\kappa,\kappa',\mathbf{q}) = \frac{1}{(m_{\mu_1}(\kappa)m_{\mu_2}(\kappa'))^{1/2}}\sum_{l'}\Phi_{\mu_1\mu_2}(l\kappa,l'\kappa')\exp[i\mathbf{q}.(\mathbf{R}(l'\kappa') - \mathbf{R}(l\kappa))] \qquad (8)$$

leads to the calculation of phonon frequencies $\omega_{qj}$ and eigenvectors $\mathbf{e}(\kappa|\mathbf{q}j)$ for values of wave vector $\mathbf{q}$ in the Brillouin zone by diagonalisation of the above dynamical matrix.

For the case of 2-D system, the motion along "c" axis, which is taken to coincide with the z- Cartesian axis is quenched . This reduces the vibrational modes to 2 per molecule. Some of the calculated phonon frequencies are presented in Table V. It was noticed that the differences in energy due to orientational repositioning of the nanotubes was very small (as also noticed from fig. 6); consequently the librational mode along 'c' axis was very low. For this very reason the mono molecule unit cell results are not different by more than 1% from the di-molecular results. Therefore we have shown the results based only on mono molecular unit cell. The phonon frequencies along (a*,0)direction is shown in fig.7. The high librational mode (One shown as curve 4) indeed is shown here despite the fact that our crystal is a 2-D crystal and a libration along, say, $a^*$ direction involves the libration of the long cylinder along a direction perpendicular to the long axis. The frequencies calculated and shown here were found to stay constant even on increasing the length of the tube. In a realistic bunch of long nanotubes, librations along directions perpendicular to long axis would have to be very low in amlitude.

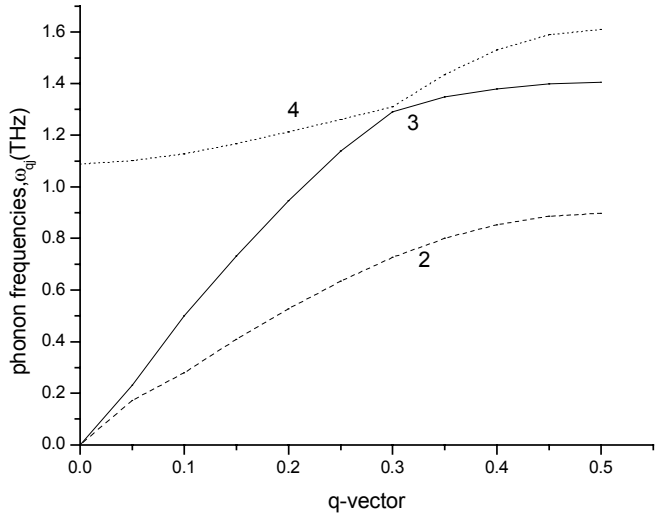

Fig 7  Phonon frequencies for a SWNT crystal along [1,0] direction. The lowest libration mode corresponding to branch 1 (not shown) has frequencies in the range .07 -.08 THz. The curves 2 and 3 refer to tanslational modes, 2 being transverse and 3 , longitudinal and curve 4 refers to librational modes along a*.

**H. Pressure Effects and Bulk Modulus**

An application of a hydrostatic pressure p alters the total potential energy such that

$$\Phi_p = \Phi + p\Delta V \qquad (9)$$

where $\Delta V$ is the increase in volume due to an application of pressure p. Therefore, a minimization of the new potential energy leads to the p-V curve.

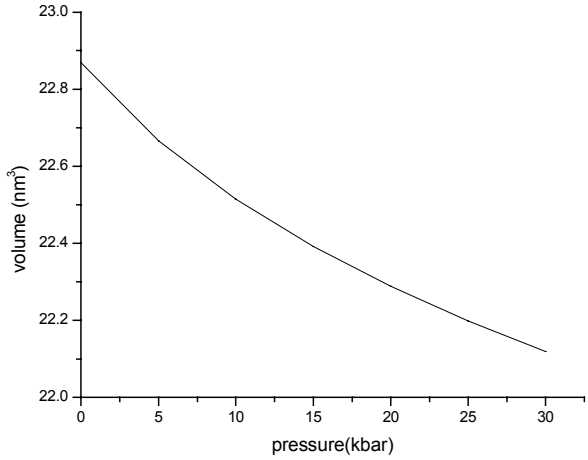

Fig. 8a

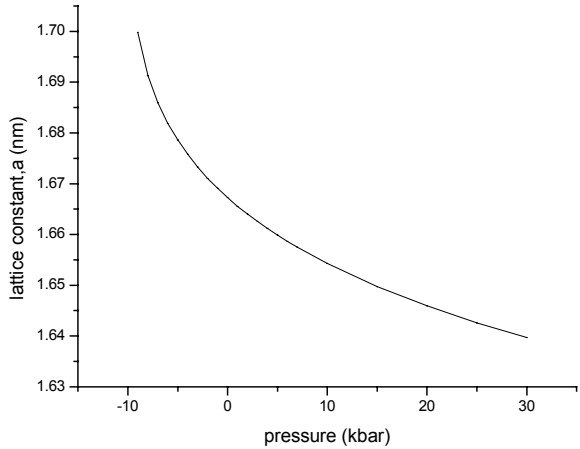

Fig. 8b

Fig 8a: The p-V curve calculated for SWNT crystal. The volume shown here corresponds to the unit cell of fixed length l along z-axis and equals √3/2 a²l. Fig 8b is just the extension of Fig 8a .The p-a curve shown here has been extended to include negative pressure values and high pressures upto 30kbar.

This data also immediately enables one to evaluate the bulk modulus, B $(= -V(\partial p/\partial V))$, and its pressure dependence. The p-V curve thus obtained is given in Fig. 8. In the case of the present model of rigid long tubes, the change in volume is basically represented by change in the curved surface area of the nanotubes. It has been found that the angle ϕ does not change on application of pressure to any significant extent for pressures from –9 kbar to 30 kbar. The bulk modulus data is presented in Table IV along with other available data. Taking 20 kbar as a reference pressure, the volume reduction comes out to be about 2.5% which is in quite good agreement with the value expected from the theortical[26] Van-der Waals compressibility of 2%. With the application of the pressure there is an increase in the volume as well as a change in the density of the material.

The energy required to compress the bunch upto 30 kbar comes out to be about 0.25 eV/C-atom (in comparison to a value of 0.18 eV/C-atom as has been measured by Chesnokov et. al.[26]).

**Table IV**

**Bulk modulus of various carbon cluster solids**

| Type of Cluster | Calculated bulk modulus (Gpa) | | Other work P=0 (Gpa) |
|---|---|---|---|
| | P=0 kbar | P=27 kbar | |
| Nanotube Crystal | 46.7 | 141.8 | 32 (Tersoff et al.[26]) |
| $C_{60}$ (Jindal et. al.[30]) | 15 | | 18 (Expt.[32]) |
| $C_{70}$ (Singh et. al.[31]) | 13 | | 11 (Expt.[33]) |

**I. Gruneisen Parameters**

Gruneisen parameters, $\gamma_{qj}$, which are related to the volume derivatives of the phonon frequencies can also be straightforwardly calculated from pressure dependence of phonon frequencies. A hydrostatic pressure leads to new volume as discussed above and new potential energy. The dynamical matrix is recalculated and pressure dependent phonon frequencies are obtained which correspond to volume dependent phonon frequencies. This leads us to calculate the mode Gruneisen parameters, defined as

$$\gamma_{qj} = -\left(\frac{\partial \ln \omega_{qj}}{\partial \ln V}\right) \quad (10)$$

The calculated values of Gruneisen parameters for some modes are presented in Table III along with mode frequencies.

Table V.

Phonon frequencies and Gruneisen constants calculated for the SWNT crystal at **q=0** and **q=(0.5,0)**

| Br.No. | Character | q-vector | $\omega_{qj}$ (THz) | $\gamma_{qj}$ |
|---|---|---|---|---|
| 1 | Librational (along c-axis) | 0.0 <br> 0.5 | 0.070 <br> 0.085 | 17.9 <br> 7.40 |
| 2 | Translational (transverse) | 0.0 <br> 0.5 | 0.0 <br> 0.897 | 0.0 <br> 11.5 |
| 3 | Translational (longitudinal) | 0.0 <br> 0.5 | 0.0 <br> 1.610 | 0.0 <br> 11.9 |
| 4 | Librational ( along a*-axis) | 0.0 <br> 0.5 | 1.088 <br> 1.406 | 11.8 <br> 11.6 |

III. **Comparison of Results and Discussion**

We now compare the numerical results for various physical quantities obtained in earlier sections for SWNT material using the simplified model based on atom-atom potentials. Firstly, a look at Fig.5 shows the potential energy of the bunch of SWNTs of unit nm length at various positions of the inter-nanotube distances. The bunch binds nicely at equilibrium distance as given in Table IIA. In this table we also present the experimental values of the lattice parameter, a, along with calculated potential energy of the SWNT bunch. This reveals that the present calculation is able to reproduce the structure very well, both for armchair and zig-zag nanotubes . The potential energy compares well with a previous calculation using cylindrical nanotube model[26]. The di-molecule unit cell, in which the neighbouring nanotubes are rotated by $180^0$ about the c-axis (Fig. 2d), does not produce any significant difference from that of mono-molecule unit cell results due to very minor difference in potential energy with respect to the orientation angle $\phi$ along c-axis. This is evidenced by looking at Fig.6 where the barrier height between the two orientational minima is also shown. The barrier height is only about 0.05 Kcal/mole/nm. Further, the periodicity with respect to orientation angle corresponds to an angle of $\phi_1 \approx 6^0$ and $\phi_2 \approx 60^0$. The angle $\phi_1$ roughly equals the angle subtended at the long axis by appropriate distance (comparable to bond length) in the hexagon of the tube surface, whereas $\phi_2$ corresponds to an overall symmetry of the unit cell. Therefore, in view of the fact that orientational fixation accounts for less than 0.1% change in the energy, for the sake of simplicity, we report the results for mono-molecule unit cell only.

We have used the model to obtain the structure of MWNT material also by calculating numerically, the diameters of the added layers which correspond to minimum energy configuration. In Table III, where the results of present calculation alongwith experimental results are tabulated, also reveal extremely good agreement for inter-wall distances. Therefore, based on our model calculations for armchair and zig-zag SWNT crystals (Table IIA), and MWNT crystals (Table III), the structures are extremely well represented by our model. The energy per C-atom (Table IIB) is comparable with other carbon cluster solids. $C_{60}$ solid however has about 30% stronger interaction energy per carbon atom.

The harmonic phonons have also been calculated by us and some of the dispersion curves for SWNT crystals have been presented in Fig. 7. Some of the external mode frequencies for this rigid molecule model corresponding to **q=0,** and q=(0.5,0) are also presented in Table V. An earlier calculation[27] also indicates that these results fall in the same range as obtained here, (5cm$^{-1}$ to 60 cm$^{-1}$). A recent calculation [34] presents interesting results on shift on intra-tube modes due to external phonons. We hope some measurements would be available soon for comparison. This table also presents Gruneisen constants which will be useful for interpreting implicit anharmonic effects.

Next, we would like to carefully look at Fig. 8b, where an equation of state curve has been plotted upto pressures of 30 Kbar. In our rigid molecule model, the p-V curve in this range shows reasonably linear behaviour. Further, the energy required to compress the bunch up to 30 Kbar of pressure, turns out to be about 0.25eV/ C-atom. On comparison with the measured value [26] of 0.18 eV/C-atom, it looks that flexibility of the nanotubes accounts for a loss of about 0.07 eV/C-atom. The bulk modulus calculated by us needs to be compared with established measured values, which at present are somewhat sketchy and disputed. (Table IV).

IV. **Summary and conclusion**

In this paper, we have attempted to present static, bulk, structural, dynamical and other phonon related properties of nanotube materials using a simple atom-atom potential for intermolecular interactions in the solid. The results indicate that some of the properties which are measured can be reproduced fairly well. The materials of SWNT have been extensively studied, and some effort has also been made to interpret MWNT diameters. The potential model that has been used here is based on C-C interactions. The same set of model potential parameters for C-C, which is part of composite set of data for C-C, C-H or D, and H-H or D-D has been used in the past to explain similar properties of a whole range of aromatic hydrocarbons. It was purposefully planned not to alter these parameters with a view to ascertaining the validity of such "universal" potentials. On the basis of results obtained here, we find that for a broad explanation of properties of these solids, simple potentials such as those provided by Kitaigorodski can be used, without making any adjustments for fullerene solids. The lattice parameters for armchair, zig-zag SWNT crystals and also for MWNT inter-wall separation have been reproduced very well. The energy per mole of the bunch of nanotubes for any reasonable length of the nanotubes, say, in the range of mm would be extremely high in comparison to any other Van-der-Waals system. Indeed, even the muliwall structures grown upto only 2 added layers for any significant length would be having large interaction energy, as can be seen from Table III. The interwall separation has come out to be nearly the same as the interplanar distance in the graphite sheets. The bulk modulus is very large (about 3 times) as compared to other carbon cluster solids, as can be noticed from Table IV. From the difference in the calculated and measured energy required to compress our crystal upto 30kbar (.25eV/C-atom and .18eV/C-atom respectively), it is evident that the nanotubes indeed at least for the diameters of around 1.2 nm are not rigid, they do undergo significant volume compression. This part of the work needs to incorporate flexibility of the nanotubes by including intra-tube interactions. Therefore, we would like to include flexible nanotubes materials for high pressure studies by including the bond bending energies and other internal tube modes, with this potential forming its basis.

**Acknowledgements:**

We gratefully acknowledge the financial support for the Research Project "Phonon Dynamics of Fullerenes and Derivatives", SP/S2/M13/96 from the Department of Science and Technology, Government of India, under which the present work was done.

**REFERENCES:**

1. H. W. Kroto, J.R. Heath, S.C.O'Brien, R.F.Curl and R.E.Smalley, Nature (London) **318**, 162 (1985). -ibid,Science **242**,1139(1986).

2. W.Kratschmer, L.D. Lamb, K. Fostiropouios, and D.R. Huffmann, Nature (London) **347**,354(1990). -ibid Chem.Phys.Lett. **17**,167(1991).

3. X. Li, J.Lu and R. M. Martin, Phys. Rev. B **46**,4301(1992).

4. E. Burgos, E.Halac andH. Bonadeo,Phys.Rev.B **47**, 13903(1993).

5. Jin Lu, Lingsong Bi, Rajiv K. Kalia and Priya Vashishta, Phys. Rev.B **49**,5008 (1994)

6. L. Pintschovius and S.L. Chaplot, Z. Phys. B **98**, 527 (1995).

7. V.K. Jindal, K. Dharamvir and Sarbpreet Singh, Int. J. Mod. Phys. B, **14**, 51 (2000).

8. J. Winter and H . Kuzmany , Physical Review B**52**,7115(1995).

7. H. Schober and B. Renker, Neutron News **10**,28 (1999).

10. Narinder Kaur, Navdeep Bajwa, K. Dharamvir and V.K. Jindal, Ind. J. Mat. Sc. 2000 -ibid Int. J. Mod. Phys. B (2000) (Submitted)

11. R.C. Haddon, A.F. Hebard, M.J. Rosseinsky, D.W. Murphy, S.J. Duclos, K.B. Lyons, B. Miller, J.M. Rosamilia, R.M. Fleming. A.R. Kortan, S.H. Glauram, A.V. Makhija, A.J. Miller, R.H. Eick, S.M. Zahurak, R. Tycko, G. Dabbagh and F.A. Thiel, Nature **350**, 320 (1991).

12. M.J. Rosseinsky, A.P. Ramirez, S.H. Glarum, D.W. Murphy, R.C. Haddon, A.F. Hebard, T.T.M. Palstra, A.R. Kortan, S.M. Zahurak and A.V. Makhija, Phys. Rev. Lett. **66**,283(1991).

13. R.M. Fleming, A.D. Ramirez, M.J. Roseinsky, D.W. Murphy, R.C. Haddon, S.M. Zahurak and A.V. Makhija, Nature **352**,787(1991).

14. P.A.Heiney et al.,Phys. Rev. Lett.**66**,2911(1991).

15. G.B. Alers, B. Golding, A.R. Kortan, R.C. Haddon and F.A. Thiel, Science **257**, 511 (1992).

16. For an overall review, see J.R.D. Copley, W.I.F. David and D.A. Neumann, Scientific Reviews of Neutron News,Vol.**l4**, No.4, Page 20(1993).

17. For a recent overview and for further refrences, see A.P. Ramirez, Condensed Matter News, Vol **3**,No. 6,Page 9(1994).

18. M.S. Dresselhaus, G. Dresselhaus and P.C. Eklund, Science of Fullerens and Carbon Nanotubes (Academic Press, Inc. 1996).

19. S. Iijima, Nature, **354** 56 (1991).

20. Boris I. Yakobson and Richard E. Smalley, American Scientist, July-August,1997.

21. For the detailed review of the properties of nanotubes, see home-page of David Tomanek's site and the site compiled by Thomas A. Adams II, http://www.pa.msu.edu/cmp/csc/ntproperties/

22. Andreas Thess, Roland Lee, Pavel Nikolaev, Hongjie Dai, Pierre Petit, Jerome Robert, Chunhui Xu, Young Hee Lee, Seong Gon Kim, Andrew G. Rinzler, Daniel T. Colbert,Gustavo Scuseria, David Tománek, John E. Fischer, and Richard E. Smalley, Science **273**, 483 (1996).

23. G. Gao, T. Cagin and W. A. Goddard, special conference issue of Nanotechnology.

Also available at site www.wag.caltech.edu/foresight/foresight_2.html

24. M.F.Yu et.al., Phys. Rev. Lett. **84**,5552 (2000)

25. J.Tersoff and R.S. Ruoff, Phys. Rev. Lett. **73**, 676 (1994).

26. S.A. Chesnokov and V.A.Nahmova, A.G.Rinzler and R.E.Smalley, J.E.Fischer, Phys. Rev. Lett. **82**, 343 (1999).

27. L. Henrard, E.Hernandez, P. Bernier and A. Rubio, Phys. Rev. B**60**, 8521 (1999).

28. K.Dharamvir, Shuchi Gupta and V.K. Jindal (to be published,2000)

27. A.I. Kitaigorodski, Molecular Crystals and Molecules (Academic Press, New York, 1973).

30. Sarbpreet Singh, K.Dharamvir and V.K.Jindal (to be published, 2000)

31. V.K. Jindal and J. Kalus, J. Phys. C **16**, 3061 (1983), and Phys. stat. sol. (b) **133**, 189 (1986).

32. S.J.Dulcos, K.Brister, R.C.Haddon, A.R.Kortan and F.A.Thiel, Nature **351**,380 (1991).

33. H. Kawamura, Y. Akahama, M. Kobayashi, H. Shinohara, H. Sato, T. Kikegawa, O. Shimomura, and K. Aoki, J.Phys. Chem. Solids **54**, 1675 (1993).

34. Daniel Kahn and Jian Ping Liu, Phys. Rev. B**60**, 6535 (1999)